# Challenges and Opportunities in Statistical Analyses and Designs for Neuroscience Clinical Trials

Authors:

Jia Jia[a]*, Man Jin[a], Bo Lu[b], Jianchang Lin[c], Inna Perevozskaya[c], Hui Yang[d], Nan Hu[e], Yi Lu[f], Qi Q[e], Jonathan Hartzel[g], Shouhao Zhou[h], Shyla Jagannatha[i], Wei Michelle Zhang[j]

**Affiliation:**

**[a]AbbVie Inc, North Chicago, IL, USA.**

**[b]The Ohio State University, Columbus, OH, USA.**

**[c]Bristol-Myers Squibb Company, Princeton, NJ, USA.**

**[d]Astellas Pharma, Northbrook, IL, USA.**

**[e]Genentech Inc., South San Francisco, CA, USA.**

**[f]Sanofi US, Cambridge, MA, USA.**

**[g]Merck & Co., Inc., Rahway, NJ, USA.**

**[h]Pennsylvania State University, University Park, PA, USA.**

**[i]Johnson & Johnson, Titusville, NJ, USA.**

**[j]Stealth BioTherapeutics, Needham, MA, USA.**

# Challenges and Opportunities in Statistical Analyses and Designs for Neuroscience Clinical Trials

## Abstract

Drug development is a complex, costly, and high-risk process, with neuroscience representing one of the most challenging therapeutic areas due to the complexity of neurodegenerative and neuropsychiatric disorders. Neuroscience clinical trials face unique barriers, making the adoption of innovative statistical methods in study design, execution, and analysis especially critical for improving clarity, feasibility, efficiency, and validity in the drug development process. The ASA Biopharmaceutical Section Neuroscience Scientific Working Group (SWG) was formed to advance statistical research and address these distinct challenges. This landscape paper provides an overview of key challenges and opportunities in statistical methods and trial designs for neuroscience, offering insights to guide future research and practice in this evolving field.

## 1. Introduction and Overview

Drug development in neuroscience is exceptionally challenging, characterized by substantial costs, extended development timelines, and significant risk. The broad spectrum of neurological and psychiatric disorders introduces distinct obstacles to clinical trial design and execution such as complex biological pathology, lack of good understanding of underlying natural disease progression, diverse and heterogeneous patient populations, pronounced placebo responses, and high dependence on subjective outcome measures. These challenges often lead to high variability in clinical data, increased rates of missing data, and greater uncertainty in interpreting trial results, all of which contribute to the elevated failure rates and unpredictability in neurological drug development.

To address these challenges, innovative statistical methods and trial design strategies are essential. Approaches such as advanced longitudinal data modeling, robust techniques for handling missing data, adaptive designs, developing biomarkers and novel endpoints, and leveraging real-world evidence can bring greater efficiency, validity, and interpretability to neuroscience clinical trials. A particular focus is needed on optimizing analysis frameworks for subjective outcomes and mitigating the impact of placebo response, which is especially pronounced in this field.

Recognizing the importance of these methodological advances, the ASA Biopharmaceutical Section convened the Neuroscience Scientific Working Group (SWG) to systematically identify and address these issues. This landscape review, authored by the SWG, presents an overview of current obstacles and promising solutions in statistical methods and trial designs relevant to neuroscience drug development. The goal is to stimulate new strategies and collaborative efforts that can accelerate therapeutic innovation for neurological disorders.

## 2. Statistical Models

Clinical trials in neurology, particularly those targeting neurodegenerative diseases such as Alzheimer’s disease (AD) and Parkinson’s disease (PD), present a unique set of analytical challenges. These disorders are typically progressive, chronic, and lack curative treatments. As a result, clinical trials often require extended follow-up duration, are susceptible to high rates of missing data, and must contend with substantial inter-subject variability in disease progression and clinical assessments.

The Mixed Model for Repeated Measures (MMRM) (Laird and Ware 1982) is the most widely used approach for analyzing continuous longitudinal outcomes in progressive diseases in neurological trials. MMRM models the mean change from baseline as a function of baseline values, categorical time points (reflecting scheduled visits), and the interaction between time and treatment. It provides unbiased estimates under the assumption of MAR and makes minimal assumptions on the correlation structure, and the shape of the outcome mean trajectory over time by choosing an unstructured correlation matrix and by treating time as categorical.

An alternative approach, Constrained Longitudinal Data Analysis (cLDA), proposed by Liang and Zeger (2000), incorporates both baseline and post-baseline outcomes within the response vector and imposes a constraint of common baseline means across treatment groups. This method includes all randomized subjects with at least one outcome observation, either baseline or postbaseline, enhances statistical power and efficiency in well-randomized trials with balanced baseline, especially when baseline and post-baseline outcomes are correlated. Both the MMRM and cLDA are accepted by regulatory agencies such as the FDA and EMA because of its minimal restrictions on the disease progression during the follow-up, although cLDA's baseline constraint is viewed with more caution.

Despite their strengths, categorical-time models like MMRM and cLDA can be overparameterized and less efficient, particularly when visit schedules are disrupted or assessments occur off-schedule - as seen during the COVID-19 pandemic or in slowly progressing diseases like AD, where functional assessments are often repeated every six months with a month window for each target date. In such contexts, modeling time as a continuous variable may offer advantages.

A constrained longitudinal data analysis with Natural Cubic Splines (NCS), as proposed by Donohue et al. (2023), provides a flexible alternative by modeling time continuously through piecewise cubic polynomials smoothly joined at predefined knots. The spline function is constructed from a set of basic functions, which are determined independently of the outcome data. This spline-based approach is more parsimonious, accommodates irregular visit schedules, and demonstrates superior power and Type I error control in simulations. It has been successfully implemented in the TRAILBLAZER-ALZ 2 (Sims et al. 2023) clinical trial for donanemab and accepted by regulatory bodies. However, NCS requires careful selection of the number and locations of the knots, assumptions about smooth temporal mean trends, which may not always align with observed data.

A limitation of MMRM and cLDA is their reliance on fixed follow-up durations, which excludes extended data from early enrollees. The Linear Mixed Effects (LME) model (Fitzmaurice et al. 2012; Chen et al. 2018) estimates treatment effects as the common slopes over continuous time for treatment and placebo groups separately. While LME offers greater power under linear progression assumptions, its applicability is limited by the rarity of truly linear or curvilinear disease trajectories (polynomial regression model) in real-world data. Consequently, LME is typically reserved for sensitivity analyses rather than primary endpoints, such as Clarity AD (van Dyck et al. 2023) clinical trial for lecanemab, which was accepted by regulatory bodies as well.

All aforementioned models accommodate both fixed effects (e.g., treatment) and random effects (e.g., subject variability), making them robust for analyzing complex or incomplete repeated measures data. The Random Coefficient Regression Model (RCRM) (Hu et al. 2021), a special case of the random effects model with time as a continuous covariate, allows for modeling individual trajectories by permitting coefficients to vary randomly across subjects. This is particularly advantageous when there is substantial inter-subject variability in response over time and offers improved model fitting and statistical power. RCRM has been applied across various research domains and regulatory settings, including FDA approvals for nintedanib, although in pulmonary fibrosis and systemic sclerosis-associated interstitial lung disease.

To address the inefficiencies of MMRM/cLDA and the linear/curvilinear assumptions of LME and RCRM, the Proportional cLDA (PcLDA) and Proportional MMRM (pMMRM) (Wang et al. 2022) have been introduced. These models estimate a consistent proportional treatment effect over time, akin to the Cox proportional hazards model. They avoid the linearity assumption of LME on disease progression, retain the categorical treatment of time but leverage all available data, including extended follow-up, enhancing power. However, their validity hinges on the proportionality assumption, which must be statistically tested. If different proportions are more appropriate at different time points, the model can be extended to include multiple proportions - though this reduces its power advantage over cLDA. In cases of moderate violations, PcLDA still yields conservative estimates. In addition, sensitivity analyses are recommended when dropout is non-random.

Bayesian Disease Progression Models (DPMs) (Wang et al. 2018; Staffaroni et al. 2022) have also emerged as powerful tools for quantifying treatment effects in neurodegenerative diseases such as autosomal dominant (AD) and frontotemporal dementia (FTD). These models define time as the estimated years from symptom onset (EYO) and model placebo trajectories as monotonic decreasing function of EYO. Treatment effect is assumed to be proportional to the decline, making the model nonlinear. While these assumptions (proportionality and monotonic decline) seem reasonable for some neurodegenerative diseases. They can be relaxed through stratification or modeling early learning effects. DPMs significantly enhance statistical power and yield clinically meaningful estimates, contingent on accurate staging or onset estimation. They have been used in studies like DIAN-TU (Bateman et al. 2017) as well as the sensitivity analysis in TRAILBLAZER-ALZ 2.

In addition to the conventional treatment effect estimated as the outcome difference at the post-baseline milestone, or the treatment effect estimated as the slope/proportion over time for two groups, Progression Models for Repeated Measures (PMRMs), proposed by Raket (2022), may offer more interpretable treatment effect metrics. This class of nonlinear mixed-effects models extends MMRM frameworks to continuous time, enabling estimation of effects such as delay or slowing of disease progression. Unlike traditional models that quantify treatment effects in outcome units (e.g., cognitive score points), PMRMs can express time-based treatment effects in time units (e.g., months of delayed decline), enhancing clinical relevance. PMRMs encompass MMRM, cLDA, and DPMs as special cases, and time-based PMRM are flexible, allowing for irregular visit schedules, multivariate outcomes, and extension to nonlinear random effects. While proportional PMRM improves power, especially when treatment effects accumulate over time, they may inflate Type I error, which can be mitigated through resampling-based calibration. PMRM has been used in the TRAILBLAZER-ALZ 2 clinical trial as one analysis included in the Type I error control strategy.

No single analytical method is universally optimal for all clinical trials in neurology. The choice of model should be guided by disease characteristics, trial design, dropout patterns, and regulatory acceptability. While MMRM and cLDA remain foundational due to their robustness and regulatory acceptance, the spline-based methods, proportional frameworks, and disease progression models offer compelling advantages. Future trials may benefit from combining flexible modeling techniques with thoughtful estimand strategies to maximize sensitivity, interpretability, and efficiency, best serve as sensitivity analyses in the plan, helping to identify the most suitable model based on trial characteristics and requirements.

An overview of commonly used methods is summarized in Appendix A.

## 3. Challenges and Opportunities in Statistical Methods

### 3.1 Placebo Effect

Placebo effect refers to the improvement in symptoms that occurs simply from receiving a substance or undergoing a procedure, even when the substance or procedure has no active therapeutic effects. The placebo effect is usually reflected as placebo response. This challenge is manifested as high failure rates of many randomized clinical trials (RCTs) with subjective outcomes. According to a systematic review paper from Furukawa et al. (2016), the average placebo response rate in antidepressant trials is about 35-40%.

The sources of placebo effects are complex and multifactorial, including treatment factors (such as expectancy-based effect, supportive contact with study clinicians), measurement factors (such as bias related to principal investigators or raters, and bias related to patients' responses), disease characteristics (such as disease subtypes of cognition), and the disease history factors (such as previous treatment experiences and outcomes). Expectancy-based mechanisms play an important role in placebo effects. Patients receiving placebo treatment may develop positive expectations or subconscious learning responses that lead to symptom improvement despite the absence of an active therapy. Interactions with clinical staff may further shape patient expectations, as healthcare providers can influence perceptions of illness and treatment, potentially improving symptoms—or, conversely, worsening them.

Beyond treatment factors, measurement factors also influence observed responses. Trials with subjective outcomes often rely on self-reports or rater assessments, which are more susceptible to patient and clinician bias than objective measures. In addition, natural disease variability can affect responses, as symptoms and severity may fluctuate independently of study procedure, further complicating the distinction between true treatment effects and placebo effects.

High placebo effect, driven by placebo responders, has significant consequences: it reduces the observable difference between treatment and placebo groups, may increase measurement error, and raise the risk of trial failure, making the development of medical products for psychiatric and other disorders time-consuming and expensive, thereby delaying the availability of potentially beneficial treatments for patients.

Mitigating the placebo effect in RCTs has several challenges. One key challenge is that while placebo responders can be observed in the placebo group, the treatment group may contain unobserved "potential placebo responders", who would respond to placebo if placebo was given. Additionally, complex data structure including baseline covariates, imaging data, biomarkers, and other factors may influence treatment response and can be difficult to control. Regulatory agencies often limit the number of covariates that can be included in an analysis, and this further complicates the cases to adjust for these confounding factors. These challenges underscore the need for more comprehensive approaches to control placebo effect and improve the reliability of treatment effect estimates in clinical trials.

Non-analytical strategies to reduce placebo effect in clinical trials include expectation management and participant training, careful site and rater selection, the use of digital phenotyping and ecological momentary assessment, genomic and neuroimaging biomarker research, and prospective blinded data analytics. These approaches seek to neutralize participant expectations, improve the accuracy of symptom reporting, and minimize variability in outcome assessments by providing standardized training and impartial communication. Selecting sites and raters with a history of low placebo response or implementing centralized training can further improve signal detection, though this may introduce logistical challenges. Digital technologies like smartphones and wearables allow for frequent, real-world symptom tracking, potentially reducing bias and identifying inconsistent responders, but require strong technological infrastructure and regulatory validation. Research into genetic and neuroimaging markers may one day help predict placebo susceptibility, although these tools are not yet practical for clinical screening. Finally, blinded data analytics support ongoing data quality monitoring and early identification of measurement error but depend on sophisticated systems and protocols. Collectively, these interventions promise to make trials more reliable, although their effectiveness and implementation challenges vary by approach and disease area.

### 3.2 **Missing Data in Neuroscience Trials**

In longitudinal clinical trials, missing data are inevitable due to intercurrent events (IEs) such as missed visits or treatment discontinuation due to lack of efficacy or safety issues. Missing data is a big challenge in neuroscience, due to higher level of missing data observed in clinical trials in neuroscience with subjective outcomes. For example, in a Phase 3 clinical trial for Parkinson's Disease (Soileau et al. 2022), there are about 8% and 37% missing data relative to safety population in control and active treatment groups. The missing mechanisms are classified into three categories (Little and Rubin 2002): Missing completely at random (MCAR) refers to the case in which the probability of the occurrence of the missing data is independent of the observed and unobserved data; missing at random (MAR) refers to the case in which the probability of the occurrence of the missing data is dependent of the observed but not the unobserved data; missing not at random (MNAR) refers to the case in which the probability of the occurrence of the missing data is dependent of the unobserved data. The commonly used statistical methods, such as linear mixed-effects model for repeated measurements (MMRM), generalized linear mixed models (GLMM), and multiple imputation (MI), are valid under MAR assumption, and they are recommended for the efficacy analysis that forms a foundation for regulatory review decision.

Since there is usually no evidence to support the MAR assumption, sensitivity analyses under MNAR assumption are often requested by regulatory agencies to assess the robustness of the MAR analysis results.

Reference-based imputation (RBI) is a commonly used strategy for MNAR. There are 3 types of RBI:

- Jump to Reference (J2R), where subjects who are prematurely discontinued are assumed to immediately lose the treatment benefit and the mean effect profile of the subjects in the active treatment group will jump to that of the subjects in the reference group after discontinuation. This method is the most conservative one among the three as it tends to remove the difference between active treatment group and the reference group instantly upon discontinuation.
- Copy Reference (CR), where the missing data in the active treatment group are assumed to have the conditional distribution after discontinuation equal to that of the reference group. This method is not as conservative as the J2R method since it still allows the carried-over treatment effect by using the values observed prior to discontinuation in the active treatment group as predictors. Thus, the mean effect profile of the subjects in the active treatment group under this assumption will gradually transition to that of the subjects in the reference group.

- Copy Increments in Reference (CIR), where the increments of the means for the active group are assumed to equal the increments of the means for the reference group at time points after discontinuation. This is the least conservative one among the three, and the most difficult to justify using in practice.

J2R, CR, and CIR were proposed for continuous endpoints a decade ago and have gained huge interest by pharmaceutical statisticians (Carpenter et al. 2013a,b; Liu and Pang 2016; Tang 2017). In recent years, Jin and Fang (2022) proposed RBI methods for repeated binary endpoints based on both multivariate probit model and logistic model, including J2R, CR and CIR.

Missing data imputation methods are widely used in clinical trials for neuroscience area when conducting a sensitivity analysis in addition to the primary analysis. For example, the CR methods were implemented in Phase 3 trial for migraine (Pozo-Rosich et al.2023) as a sensitivity analysis.

Return to Baseline (RTB) is another method to handle MNAR data by imputing the missing data using baseline values, assuming the responses of subjects who prematurely discontinue the study treatment would lose the treatment benefit quickly and “return to baseline”. RTB may work in some disease areas, but it is not a recommended strategy for neurodegenerative diseases, such as Alzheimer’s Disease, which progress over time regardless of the status of treatment. RTB would overestimate the post-baseline measurements by using baseline measurement.

### 3.3 **Subjective Outcome Measurements**

One of the most common challenges in neuroscience is subjective outcome measures. Unlike objective measures which are immune to human factor and thus more easily quantified, subjective measures are usually patient-reported outcomes (PRO) which rely on patients’ description of personal experience and feelings on a specific condition such as quality of life.

A more generalized type of outcome measures of interest, though, doesn’t appear to be subjective as measured by a trained Investigator or rater, but not completely objective either, as the results could be impacted by patients’ expectations and efforts. This could lead to the issue of placebo effect and even large variation of the placebo or standard of care (SoC) effect in controlled clinical studies, depending on the setting of each study.

For example, the Medical Research Council (MRC) scale and grip strength are tools used to evaluate muscle strength, particularly in cases of peripheral neuropathy like Guillain-Barré syndrome (GBS) and chronic inflammatory demyelinating polyradiculoneuropathy (CIDP). For both assessments, the patient's effort is measured by a qualified rater. As shown in Table 1, when comparing immunoglobulin (Ig, considered as one of the current SOCs for CIDP) with placebo in CIDP patients, in the ICE study (Hughes et al. 2008) with SoC-naïve population, both arms demonstrated positive trend on change from baseline in MRC scale and grip strength during first period after randomization. However, in the PATH study (van Schaik et al.2018) with randomized withdrawal design in CIDP patients who responded and stabilized on intravenous (IV) Ig treatment before randomization, all treatment groups showed negative trend in change from baseline in MRC-SS and grip strength, after withdrawal from IVIg (placebo group) or switching to subcutaneous (SC) Ig. The different performance of Ig and placebo in MRC scale and grip strength as observed may reflect patients' expectation of treatment effect under different context, e.g. generally positive if possibly on new promising treatment, while non-worsening at best if possibly withdrew from SoC. Furthermore, the outcome measure itself, if depending on the patients' behavior, could allow such expectation to impact the results via patients' efforts proactively or subconsciously. Most importantly, it's hard to quantify the impact of patients' subjectivity factor on top of the actual treatment effect under the context of each study.

Table 1. Change from baseline in MRC scale and grip strength in ICE study and PATH study

| | **ICE study first period Mean (SD)** | | **PATH study Median (IQR)** | | |
|---|---|---|---|---|---|
| | **Placebo (N=58)** | **IVIg (N=59)** | **Placebo (N=57)** | **Low-dose SCIg (N=57)** | **High-dose SCIg (N=58)** |
| **MRC sum score** | 0.2 (4.5) | 3.3 (5.6) | -2.0 (-6.0 to 0.0) | 0.0 (-2.0 to 2.0) | 0.0 (-2.0 to 1.0) |
| **Grip strength, dominant hand (kPa)** | 1.5 (15.6) | 13.2 (19.3) | -6.6 (-21.6 to 0.3) | -0.6 (-8.9 to 7.0) | -2.7 (-6.6 to 2.0) |

The variability of subjective outcome measures with the same treatment across different contexts leads to issues of comparability. Within a study, even in randomized studies with concurrent control, it's unclear whether such subjectivity factor affects the treatment of interest and the control (placebo or SoC) to the same extent. Moreover, through studies, in new trial design where assumptions are based on historical data, or concurrent control is limited (e.g. in rare disease indication) such that historical control is borrowed, the uncertainty of the subjective outcome measures could highly increase the risk in trial cost (sample size) and probability of success (power). To mitigate such risk, meta-analysis on systematic review of historical data may be considered to more precisely depict average placebo or SoC effect by different settings.

### 3.4 Less-Than-Expected Signal-To-Noise Ratio

The primary endpoint in neuroscience Phase 3 clinical trials is traditionally chosen to be a continuous endpoint, e.g. change from baseline in CDR-SB (van Dyck et al. 2023), annualized rate of change in FCSRT CI (Tariot et al. 2026). To ensure adequate power for the Phase 3 studies, signal-to-noise ratio of the continuous endpoint needs to be assumed.

An examination of recently completed Autosomal-Dominant Alzheimer's Disease (ADAD) trials revealed concerns about the choice of the primary endpoint, especially the signal-to-noise ratio assumed at the trial design stage (Mackey et al. 2021). The authors found that placebo groups did not progress in the expected direction in some ADAD trials, while some yielded a smaller than expected signal-to-noise ratio. The placebo effect could be a major reason contributing to these findings. Other reasons include the potential trial population shift from the assumed population and other unknowable factors.

We invite trial practitioners in neuroscience to caution the less than expected signal-to-noise ratio, especially for endpoints that have not been validated in a clinical trial setting. Some recent publications aim to safeguard this issue. Hu et al. (2025) provides a Type I error controlled approach that adaptively chooses a primary endpoint based on blinded data. Xu et al. (2023) offers a Type I error controlled approach to unblind part of the study to choose a primary endpoint. In general, more research is welcomed in this area such as building good predictive models for natural disease progression to better understand and model disease progressions, which may help to make more realistic assumptions about expected noise to signal ratio when planning clinical trials. Different endpoints may need to be utilized in early and late-stage development and building a good translation model quantifying relationships between early and late-stage endpoints may help with right-sizing the clinical trials in neuroscience trials.

### 3.5 Sample Size Reduction in Neuroscience Trials

The heterogeneous symptoms and varying severities of neurological conditions frequently introduce significant variability into clinical trial data, complicating efforts to reduce sample size. Furthermore, since long-term follow-up is often required in neuroscience trials to assess the durability of treatment effects, small sample sizes exacerbate issues related to participant dropout and missing data, making analysis and interpretation more challenging. Consequently, reducing sample size in longitudinal studies remains a substantial challenge.

Random coefficient regression models (RCRM) (Hu et al. 2021) can offer superior power and modeling flexibility in certain situations compared to MMRM. MMRM is widely used due to its robustness in handling missing data. By accounting for both fixed effects (e.g., treatment) and random effects (e.g., subject variability), MMRM effectively manages complex repeated measures data, especially when data is incomplete or unbalanced.

RCRM enables modeling of individual trajectories by allowing coefficients to vary randomly across subjects. This is particularly advantageous when there is significant inter-subject variability, as it accommodates personalized patterns of change over time. By explicitly capturing individual differences, RCRM enhances the power to detect subject-specific effects, which would otherwise be diluted in models assuming constant effects.

Furthermore, RCRM can model nonlinear relationships between time and response variables by incorporating random coefficients for polynomial terms, which is a flexibility often limited in standard MMRM frameworks. This adaptability allows RCRM to better capture complex temporal dynamics, improving model fit and potentially increasing statistical power. Ultimately, the increased flexibility and statistical power of RCRM in handling heterogeneity and non-linear trends can translate into smaller required sample sizes.

Similar to RCRM, researchers proposed natural cubic splines model (Donohue et al. 2023) to also treat time as continuous, however, disease trajectory follows a flexible natural cubic spline. This model is shown to improve power compared to the traditional MMRM. This method has been pre-specified in a Phase 3 Alzheimer's clinical trial (Sims et al. 2023).

Another model that was proposed recently is called the proportional MMRM (pMMRM) (Wang et al. 2022). While this model does not assume linearity, it treats time variable as categorical and assumes the treatment effect is consistent across all visits. The authors show that the pMMRM can lead to power gain up to 20% to 30%.

## 4. Challenges and Opportunities in Trial Designs

Clinical trial designs in the neuroscience therapeutic area are largely similar to those used in other fields. However, given the unique challenges of neurological diseases—such as low prevalence in many neurologic disorders, slow progression, high variability in patient responses, and long-term follow-up—it can be advantageous to increase utilization of innovative trial designs specifically tailored for neuroscience studies. Adopting such approaches may help enhance trial efficiency, improve data quality, expedite or de-risk development and yield more clinically meaningful results.

### 4.1 Designs to Leveraging External Trial and Real-World Data

One major limiting factor of running conventional randomized clinical trials for neurological diseases is the difficulties in recruitment due to rarity of the disease. For example, amyotrophic lateral sclerosis (ALS) and Duchenne muscular dystrophy (DMD) are two rare diseases with an incidence of 2 cases per 100,000 population in major Western countries for ALS and 2.8 cases per 100,000 population for DMD (Stephenson et al. 2022). But both are devastating chronic progressive diseases involving neuromuscular system structure and function and are in urgent need for effective therapies. The relatively slow progress in treatment development calls for change and innovation in trial design and the approval process of regulatory agencies. Recommended approaches included relaxing the requirements of placebo arms from clinical trials and taking advantage of external control patients from real-world data (RWD, Stephenson et al. 2022).

Formally, RWD are defined as data relating to patient health status or the delivery of health care routinely collected from a variety of non-RCT sources (U.S. Food and Drug Administration 2021a). The benefits of using RWD are multifold: RWD may better reflect actual clinical settings in which interventions are applied, hence to help refine eligibility criteria to better represent actual patient populations; RWD can be used to augment underpowered control groups or provide external validation; Longitudinal RWD may offer insight into long-term treatment effects beyond trial duration, which is particularly important for many neurological diseases. On the other hand, the proper use of RWD also faces challenges, from relevance/quality of external data and trial integrity and transparency, to result interpretability.

From a statistical perspective, valid causal inference involving non-randomized components often requires assumptions on consistency (between observed and potential outcomes), exchangeability (of treatment assignment) and positivity of treatment assignment (Little and Rubin 2000). For leveraging external control data to supplement clinical trials, we also need an exchangeability assumption on data source (Valancius et al. 2024). It literally states the conditional independence (given observed covariates) of potential outcomes under control between the trial population and the external patient population.

Data borrowing can be implemented from multiple sources, including real-world data (RWD), as exemplified in the aforementioned ALS study, or from historical clinical trial datasets to supplement data from ongoing trials. In the latter scenario, regulatory authorities may exhibit a greater willingness to accept Bayesian methodologies for data borrowing, given the superior data quality typically associated with clinical trials relative to RWD. Borrowing may be restricted to the control arm—representing a less complex approach—or extended to both treatment and control arms, which presents greater methodological challenges. In both cases, it is imperative that patient characteristics between the historical and current trial populations are comparable to ensure the validity of the exchangeability assumption. Pragmatic strategies for establishing exchangeability, as well as recommendations for implementation, are elaborated in recent draft guidance documents from the FDA regarding Bayesian designs and from the EMA.

Assuming exchangeability, statistical inference can be performed using various approaches. Commonly employed techniques include Bayesian data borrowing methods (such as the power prior) and frequentist approaches, including propensity score-based methods (matching, weighting, and stratification) (Fu et al. 2023). At present, there is no definitive guidance regarding the circumstances under which each method should be utilized; the selection depends on factors such as data availability, data quality, and trial objectives. The key concepts underlying these methods are summarized below. Although not specific to neuroscience, they are particularly valuable in contexts characterized by recruitment challenges in rare diseases (e.g., ALS) or by limited historical data on the natural history of diseases such as Alzheimer's disease.

**Bayesian Borrowing Method**

Bayesian methods are a natural way to incorporate external data information to assist the inference. These approaches encode external evidence into the prior distribution for the treatment effect parameter and then modulate the contribution of this prior based on concordance with observed trial data.

The common trait of these methods is using informative priors (as opposed to non-informative priors that are intended to be diffuse and not informed by historical data). Informative priors used in methods below incorporate information from external data into the prior for the current trial parameter. The posterior distribution under borrowing replaces the non-informed prior with an externally informed prior, so that external information effectively enters the likelihood-driven update mechanism. There are three most commonly used approaches that are well established and may be useful in neuroscience to expedite clinical development because they allow to increase precision of trial without drastically increasing resources: power prior, robust meta-analytic prior, and commensurate prior.

**Power Prior**

Power prior includes the likelihood function of the external data as part of the prior for the parameter of interest, e.g. the mean outcome or the mortality rate from control patients (Ibrahim et al. 2015). This external data likelihood component is raised to the power of $\alpha$, where $\alpha$ is the power parameter that weighs the external data relative to the likelihood of the current data based on similarity, with $0 \leq \alpha \leq 1$. $\alpha = 0$ indicates that no external data is borrowed, while $\alpha = 1$ implies that external data contribute equally as the current data to inference. Usually, $0 < \alpha < 1$ is determined by the distribution overlap between the current and external data. The power prior represents a static approach to borrowing external data, as the borrowing parameter $\alpha$ must be specified a priori and is not determined from the data itself. If $\alpha$ is misspecified, there is a risk of introducing bias and inflating Type I error rates. This limitation is a well-recognized critique of power prior methods, and subsequent work by Ibrahim and colleagues has introduced several methodological extensions to mitigate these concerns.

**Robust Meta-Analytic Prior**

Robust meta-analytic prior (RMAP), in contrast, is a dynamic borrowing method where amount of borrowing from historic data is driven by the degree of concordance/discordance between the two data sources (current and historic). The method is built on the concept of Meta-Analytic Predictive (MAP) Prior (Neuenschwander et al. 2010), which constructs a hierarchical model over multiple historical studies, assuming study-specific effects arise from a common distribution with mean $\mu$ and heterogeneity parameter $\tau$. ($\tau \rightarrow \infty$ implies complete pooling (identical studies) and $\tau \rightarrow 0$ implies no borrowing). The Robust MAP prior (RMAP) (Schmidli et al. 2014) extends MAP by mixing an informative MAP component with a non-informative ("robust") component. The mixture weight ω in the robust meta-analytic prior (RMAP) framework represents initial skepticism about historical data. As posterior updating occurs, the external information is naturally down-weighted when prior–data conflict is detected, thereby improving robustness to drift or heterogeneity. RMAP thus allows dynamic modulation of borrowing strength and provides heavier tails to mitigate misleading historical information.

**Propensity Score**

**Matching**

Matching is a nonparametric way to balance covariate distribution between treated and control groups, based on a pre-specified distance metric. Matching on propensity scores resembles a block randomized design, which is easily interpretable to lay audience. Propensity scores are defined as the conditional probability of receiving treatments given observed covariates (Lu 2021). The main advantage is that the matching process does not involve the outcome, and the statistical inference is conducted separately after matching. Different matching strategies can be implemented depending on the original study design, hybrid randomized trial or single-arm trial (Liu et al. 2022). Only a portion of external data, the control patients who are mostly like the current population, will be used to ensure the exchangeability. Matching design picks high-quality control to supplement the trial. It is considered as partial borrowing as it does not need to use all external data.

**Weighting**

Weighting is another popular way for propensity score adjustment. Treated and controlled patients are weighed differently to create a pseudo-population in which all covariates are balanced to remove confounding biases. Usually, treated patients are weighed by the inverse of the propensity score and control patients are weighed by the inverse of one-minus-the-propensity score. Weighing tends to use all external control data except in scenarios where controls with extreme weights need to be trimmed. One additional advantage of weighting is to combine it with outcome regression modeling to create a doubly robust estimator (Bang 2005). Such estimator can obtain consistent treatment effect estimation when either the propensity score model or the outcome model is correctly specified. Weighting methods usually use all external data so it is considered as full borrowing.

**Stratification**

Stratification groups patients by similar propensity score values into several strata. Stratum-level causal estimation is first conducted; the final estimate is then combined across all strata. It is easier to implement than matching or weighting. It may be combined with Bayesian power prior method to further reduce the bias within each propensity score stratum (Wang et al. 2019).

Though those methods are well established to handle observational data in causal inference, their implementation for external control borrowing in practice still faces numerous challenges.

Firstly, external data quality is crucial for establishing causal evidence. High quality data are more likely to satisfy the exchangeability assumption, hence, to warrant valid causal interpretation (U.S. Food and Drug Administration 2021b). Historical control data from recent clinical trials on the same disease may be a good source, as they need to have similar inclusion/exclusion criteria to ensure comparability. For chronic neurological diseases, natural history data could be a valuable source in understanding disease progression for patients not undergoing the treatment (Li and Izem 2022). Secondly, the Bayesian power prior method uses a power parameter to discount the external data which is considered not that similar to the current population. But such discounting treats every control equally, not distinguishing between good and poor controls. This is not a very efficient use of external data. Qian et al. (2025) proposed a matching-assisted power prior method to first select comparable control data via matching, then incorporate them into the analysis by using power prior, which showed improved performance. Thirdly, the major threat to the exchangeability assumption is the existence of unmeasured confounding, implying that the trial population and the external control patients may differ in ways not captured by the observed covariates. Exchangeability is a convenient but unverifiable assumption. More rigorous causal evaluation needs to consider sensitivity analyses to assess the impact of potential unmeasured confounding (Lu 2021; Sjölander et al. 2022). In many neurological disease studies, the endpoints are functional outcomes at later time points. Liu et al. (2022) introduced the idea of intermediate outcome assisted borrowing, which takes advantage of the intermediate outcomes measured earlier in the study to control unmeasured confounding.

### 4.2 Platform Trials

Traditional (fixed) clinical designs are slow, expensive, and carrying high risk of failure in all TAs, inefficient due to slow progression of neurodegenerative or neurodevelopmental disorders, long follow-up times, heterogeneous patient populations, multiple mechanisms of disease, pronounced placebo effects, and modest signal-to-noise ratio. Adaptive trials and, particularly platform trials, can offer a viable solution to alleviate these challenges.

Aiming to accelerate therapeutic development, platform trials have emerged as a transformative paradigm for clinical development (Park et al. 2019; Gittings et al. 2020). At its core, a platform trial allows multiple interventions to be tested simultaneously (and sequentially) in a master protocol with a shared control arm and adaptive decision-making rules (Saville et al. 2016; Koenig et al. 2023). By integrating adaptive randomization, shared control groups, and pre-specified decision rules, platform trials facilitate resource efficiency while maintaining rigorous statistical control.

Since multiple therapies are required to be studied in one trial, the initial use of platform trials was most prominent in disease areas where a lot of drugs are emerging rapidly for testing, such as oncology. A well-known example of a platform trial form oncology is I-Spy 2 for breast cancer (Pignatti et al. 2015). However, as experience and level of comfort with them grew and sources of efficiency became more apparent, the use of platform trials expanded to other therapeutic areas such as rare diseases where patient scarcity presents its own set of challenges to study multiple therapies simultaneously. In disorders such as ALS, Alzheimer's disease, Parkinson's disease, or autism spectrum disorder, the promise of such an infrastructure is particularly strong (Aisen et al. 2021; Paganoni et al. 2022; Wathen et al. 2023; Fabbri et al. 2024). Examples of such application of platform trials in NS include HEALEY amyotrophic lateral sclerosis (ALS) Platform Trial, which demonstrated great potential by reducing the cost of research by 30%, decreasing the trial time by 50%, and increasing patient participation by 67%, to accelerate evidence generation in clinical neuroscience practice (Paganoni et al. 2022).

In particular, continuous evaluation of multiple therapies in a platform trial is implemented with adaptive entry/exit of treatments. Unlike a typical multi-arm trial conducted only once, or a series of independent two-arm trials, a platform design is "perpetual" or open-ended: as new therapeutic candidates emerge, they can be integrated into the ongoing infrastructure. Moreover, the shared-control design reduces the need for repetitive control arms and the overhead of repeated trial infrastructure, making it attractive to patients with unmet medical needs such as ALS or Alzheimer's. In neuroscience, where heterogeneity in progression, endpoints, and mechanisms is the norm, the flexibility and scalability of the platform design offer tangible advantages. Also, having a concurrent control arm run for a long time allows researchers to gain insight into natural disease progression, which often is a challenging problem in neuroscience. Platform trials are seldom designed for a particular duration but to serve as a potentially perpetual structure over which sequential trials can be developed (Koenig et al. 2023). So, besides boosting current trial efficiency, concurrent arm in neuroscience platform trials can provide valuable insights helping design future trials.

Importantly, due to multiple and complex decisions to be made (early stopping, changing allocation ratio, etc.), the statistical framework often utilizes Bayesian designs with interim analyses that may include borrowing of information across arms and longitudinal modelling across time. Response-adaptive randomization, which dynamically updates posterior probabilities of efficacy or futility as new data accrue, is also often used (Pitre et al. 2023; Massonnaud et al. 2025). To preserve overall Type I error under such flexible adaptations, advanced design planning and calibration strategies, such as multiplicity-adjusted decision rules, concurrent control restrictions, and simulation-based operating-characteristic evaluations, are typically required to ensure both long-run frequentist validity and Bayesian adaptability, ultimately facilitating regulatory transparency.

While challenging, platform trial designs also unlock several key opportunities for neuroscience research. First, the ability to integrate external or historical controls into the shared control arm can further reduce the number of placebo assignments, which is especially valuable in rare neuro-disorders or pediatric populations (Bofill Roig et al. 2023). Second, more efficient algorithms of adaptive randomization and early stopping rules can enable assignment weights to shift toward more promising arms, reducing exposure to ineffective therapies, focusing resources, and making it more attractive to patients (Robertson et al. 2023). Third, the synergy with emerging digital biomarkers, wearable sensor data, and artificial intelligence–driven interim analyses offers unprecedented agility: continuous data streams (e.g., gait sensors in PD, speech markers in ALS) can trigger adaptive decisions faster than traditional periodic assessments (Johnson et al. 2023). Fourth, the platform design can facilitate seamless phase 2 to phase 3 transitions, which is especially relevant in neuroscience where early signals might be derived from imaging/biomarker surrogates (Park et al. 2019) but the relationship between surrogates and clinical (long-term) endpoint needs to be validated.

Looking ahead, we anticipate that neuroscience research will increasingly adopt collaborative, multi-sponsor platform trial infrastructures. To support this transition, developing shared infrastructure (data coordinating centers, sensor networks, master protocols) and building training and capacity for Bayesian adaptive analytics will be essential. For neuroscience investigators, early planning of platform architecture in governance, data management, endpoint harmonization, and efficient adaptive rule development can lead to significant long-term dividends in efficiency, speed, and learning. Overall, platform trials represent a paradigm shift in how we design neuroscience therapeutic trials, bringing the opportunity to help accelerate the translation of neurotherapeutics from bench to bedside when testing many more therapies faster and more ethically, yet demanding rigorous statistical design, operational infrastructure, and regulatory foresight.

### 4.3 **Designs Approaches For Managing Placebo Response**

Section 3.1 discussed the source and impact of the placebo response to clinical trials along with several non-analytical mitigation strategies. Several study design strategies have also been applied, one of the earliest involving placebo lead-in periods, where participants receive placebo during a run-in phase before randomization. Those showing early improvement (presumed placebo responders) are then excluded from the trial. The rationale was straightforward: remove individuals most susceptible to placebo effects before randomization, thereby enriching the study population with true drug responders. This approach offers practical advantages, such as being easy to incorporate into current clinical trial frameworks, potentially identifying early responders who could confound results, and providing a way to assess participant adherence and commitment. However, its effectiveness has not been rigorously validated in randomized controlled trials. There are concerns that it may unintentionally exclude patients who would benefit from the investigational treatment, and it can lead to longer trial durations and higher costs without proven improvements in study outcomes. Meta-analyses examining placebo lead-in effectiveness have concluded that this approach does not lower the placebo response rate (Trivedi et al. 1994) and is no longer recommended as a primary mitigation strategy.

Another approach is Sequential Parallel Comparison Design (SPCD) which conducts the trial in two sequential steps (Fava et al. 2003). In step 1, all participants are randomized to drug or placebo. In step 2, placebo non-responders from step 1 are re-randomized to drug or placebo, while drug responders continue drug. This design theoretically enriches step 2 with placebo non-responders, improving signal detection. This design offers several advantages, including increased statistical power by leveraging placebo non-responders, the potential to reduce sample size compared to traditional approaches, and multiple opportunities to detect treatment effects throughout the trial (Tamura et al. 2007). However, it comes with notable drawbacks: the design is complex and requires specialized statistical analysis, the trial duration is considerably longer due to sequential phases, and dropout rates may rise as participants are required to remain in the study for extended periods. Additionally, regulatory acceptance can vary and necessitate comprehensive justification and planning, while ethical concerns may arise from prolonging exposure to placebo for some trial participants. While SPCD has shown some promise in specific contexts, due to its limitations on clinical interpretability and statistical analysis it has not been used in confirmatory trial setting (Tsuchida et al. 2021).

### 4.4 **Adaptive Designs**

Adaptive clinical trial designs are increasingly recognized as a flexible and innovative strategy to improve efficiency in neuroscience drug development, particularly in settings characterized by substantial uncertainties in effect size, disease progression, and patient heterogeneity. Unlike traditional fixed designs, adaptive trials allow for prospectively planned modifications, such as sample size re-estimation, dose selection, population enrichment, or dropping ineffective arms, based on interim data, while maintaining trial integrity and statistical validity (U.S. Food and Drug Administration 2019a; Chow & Chang 2011).

In neurodegenerative and neuropsychiatric disorders, adaptive designs offer distinct advantages. These approaches enable early learning and dynamic decision-making in diseases characterized by slowly evolving and heterogeneous progression such as Alzheimer’s disease and Parkinson’s psychosis. They may also mitigate risks associated with high placebo response rates and substantial inter-patient variability (Berry et al. 2010; Cummings et al. 2014).

The development program for Aduhelm (aducanumab) provides an example of adaptive interim monitoring in neurodegenerative disease. The pivotal Phase 3 EMERGE and ENGAGE trials incorporated a prespecified interim futility analysis, a group-sequential adaptive feature allowing early termination based on accumulating efficacy data. Although the trials were initially stopped for futility, continued follow-up and additional analyses contributed to the U.S. Food and Drug Administration (FDA) accelerated approval decision. This case highlights the regulatory relevance and complexity of adaptive interim analyses when prospectively planned and appropriately interpreted (Budd Haeberlein et al. 2022).

Bayesian adaptive methodologies have been further advanced in the Alzheimer's disease program for Leqembi (lecanemab). The Phase 2b BAN2401 study employed a Bayesian adaptive design with response-adaptive randomization, enabling dynamic allocation across multiple dose levels and dosing regimens based on posterior probabilities of treatment benefit. This approach supported efficient dose-regimen learning and informed the confirmatory Phase 3 development program, demonstrating how Bayesian adaptive designs can enhance efficiency, ethical treatment allocation, and regulatory decision-making in early-stage neurodegenerative trials (Satlin et al. 2016; Swanson et al. 2021; U.S. Food and Drug Administration 2026).

In neuropsychiatric drug development, clinical programs apply specialized adaptive framework. One such design is the Sequential Parallel Comparison Design (SPCD), a two-stage adaptive design developed to address high placebo response in major depressive disorders. SPCD adaptively re-randomizes placebo non-responders in the second stage and combines evidence across stages, improving statistical power and efficiency while maintaining strict control of the type I error rate (Papakostas et al. 2009; U.S. Food and Drug Administration 2019b).

Despite these advantages, the implementation of adaptive designs in neuroscience presents important challenges. Long trial durations, delayed treatment effects, and complex endpoints, such as cognitive, functional, and behavioral composites, increase the risk of operational bias and complicate interim assessments. Reliance on surrogate endpoints or subjective rating scales may further limit the robustness of adaptation decisions. Consequently, adaptive trials in neuroscience require extensive simulation work, clearly prespecified adaptation rules, and early, ongoing regulatory interaction to ensure control of type I error and interpretability of results (U.S. Food and Drug Administration 2020; Pallmann et al. 2018). Operational execution, including maintenance of blinding, data quality, and trial consistency, is particularly demanding in large, global, multicenter studies.

Nevertheless, adaptive designs represent a promising pathway to enhance feasibility, efficiency, and decision-making in neuroscience clinical development. Continued collaboration among statisticians, clinicians, regulators, and data monitoring committees will be essential to optimize their implementation. Further methodological guidance, particularly on Bayesian adaptive frameworks, the integration of digital and biomarker endpoints, and alignment of adaptive strategies with clinically meaningful estimands will be critical to advancing the responsible application of adaptive designs in neuroscience trials.

## 5. Summary

In summary, neuroscience drug development presents exceptional methodological and operational challenges, including heterogeneous disease progression, subjective outcome assessments, pronounced placebo effects, high rates of missing data, and substantial uncertainty in endpoint behavior and treatment effect size. This review highlights that while traditional methods such as MMRM and cLDA remain important foundations, emerging approaches—including spline-based longitudinal models, proportional treatment-effect models, disease progression models, reference-based imputation strategies, and adaptive endpoint-selection methods—offer meaningful opportunities to improve efficiency, interpretability, and robustness. At the design level, innovative strategies such as external control borrowing, platform trials, placebo-mitigating designs, and broader adaptive frameworks have strong potential to address the unique constraints of neuroscience trials, particularly in rare, slowly progressive, and highly variable diseases. Continued progress will depend on close collaboration among statisticians, clinicians, regulators, and industry to align methodological innovation with practical implementation and regulatory expectations. By advancing both statistical methodology and trial design infrastructure, the neuroscience community can strengthen evidence generation and help accelerate the development of safe and effective therapies for patients with neurologic and psychiatric disorders.

**Disclosure Statement**

The authors report there are no competing interests to declare.

Appendix A

**Table A1: Overview of Longitudinal Data Analytic Approaches in Neurology Clinical Trials**

| | | |
|---|---|---|
| MMRM | Assumptions | • MAR assumption<br>• minimal mean trajectory<br>• time as categorical |
| | Estimation | mean difference between arms at specific timepoint |
| | Strength | • unbiased estimates under MAR<br>• minimal model assumptions<br>• widely accepted by regulatory |
| | Limitation | • overparameterized<br>• less efficient with irregular schedules |
| | Best use | progressive diseases with regular assessments |
| | Example trials | general use in neurology trials |
| cLDA | Assumptions | • MAR assumption<br>• common baseline means across arms<br>• time as categorical |
| | Estimation | mean difference between arms at specific timepoint |
| | Strength | • include all randomized subjects<br>• efficient in balanced randomization |
| | Limitation | • sensitive to baseline imbalance<br>• assume common baseline value across arms |
| | Best use | randomized trials with baseline balanced |
| | Example trials | general use in neurology trials |
| NCS | Assumptions | • smooth temporal mean trends<br>• pre-defined knots<br>• time as continuous |
| | Estimation | mean difference between arms at specific timepoint |
| | Strength | • flexible and more parsimonious<br>• accommodates irregular schedules<br>• superior in power and Type I error control |
| | Limitation | • requires careful selection of number and location of knots<br>• assumptions about smooth trends |
| | Best use | • trials with large sample size and more visits<br>• irregular visit schedules |
| | Example trials | TRAILBLAZER-ALZ 2 |
| LME (slope analysis) | Assumptions | linear or curvilinear trajectory |
| | Estimation | common slopes over time (rate of change) |
| | Strength | • greater power under linear assumptions<br>• simple interpretation |
| | Limitation | • limited by rarity of linear trajectories<br>• poor fit with nonlinear progression |
| | Best use | • sensitivity analyses<br>• long term follow up |
| | Example trials | Clarity AD |
| RCRM | Assumptions | Substantial inter-subject variability in response over time |
| | Estimation | model individual trajectories with random slopes and intercepts |
| | Strength | improved model fitting accommodates inter-subject variability |
| | Limitation | limited by linear/curvilinear assumptions |
| | Best use | substantial inter-subject variability in response over time<br>long term follow up |
| | Example trials | FDA approvals for nintedanib; AD prevention trial testing efficacy and safety of crenezumab |
| PcLDA/ pMMRM | Assumptions | proportional treatment effect over time |
| | Estimation | constant proportional treatment effect in decline |
| | Strength | • avoids linearity assumption<br>• leverages extended data<br>• more efficient |
| | Limitation | • validity hinges on proportionality assumption<br>• loss of power if violated |
| | Best use | modest violations of proportionality |
| | Example trials | n/a |
| DPMs | Assumptions | • proportionality<br>• monotonic decline |
| | Estimation | nonlinear disease modeling to see treatment difference |
| | Strength | • enhanced power<br>• clinically meaningful estimates<br>• models progression vs symptom onset (EYO) |
| | Limitation | requires accurate staging or onset estimation |
| | Best use | preclinical or early-onset neurodegeneration |
| | Example trials | • DIAN-TU<br>• TRAILBLAZER-ALZ 2 |
| PMRMs | Assumptions | • flexible<br>• nonlinear mixed-effects<br>• potential proportional effect |
| | Estimation | time-based effects (delay/slowing) in time units |
| | Strength | • intuitive interpretation (e.g., delay in months)<br>• accommodates irregular visits<br>• generalizes MMRM/cLDA/DPM |
| | Limitation | may inflate Type I error, need calibration |
| | Best use | • regular schedules<br>• multivariate outcomes<br>• flexible sensitivity/supplementary analysis |
| | Example trials | TRAILBLAZER-ALZ 2<br>(primary analysis included in Type I error control framework) |